\documentclass[doublecol]{epl2} 

\usepackage{amsmath,amssymb,upgreek,graphbox}

\title{Velocimetry in rapidly rotating convection: spatial correlations, flow structures and length scales}
\shorttitle{Velocimetry in rapidly rotating convection}

\author{Matteo Madonia\inst{1} \and Andr\'es J. Aguirre Guzm\'an\inst{1} \and Herman J. H. Clercx\inst{1} \and Rudie P. J. Kunnen\inst{1}\thanks{E-mail: \email{r.p.j.kunnen@tue.nl}}}
\shortauthor{M. Madonia \etal}

\institute{                    
  \inst{1} Fluids and Flows group, Department of Applied Physics and J. M. Burgers Centre for Fluid Dynamics, Eindhoven University of Technology, P.O. Box 513, 5600 MB Eindhoven, The Netherlands, EU
}


\abstract{Rotating Rayleigh--B\'enard convection is an oft-employed model system to evaluate the interplay of buoyant forcing and Coriolis forces due to rotation, an eminently relevant interaction of dynamical effects found in many geophysical and astrophysical flows. These flows display extreme values of the governing parameters: large Rayleigh numbers $Ra$, quantifying the strength of thermal forcing, and small Ekman numbers $E$, a parameter inversely proportional to the rotation rate. This leads to the dominant geostrophic balance of forces in the flow between pressure gradient and Coriolis force. The so-called geostrophic regime of rotating convection is difficult to study with laboratory experiments and numerical simulations given the requirements to attain simultaneously large $Ra$ values and small values of $E$. Here, we use flow measurements using stereoscopic particle image velocimetry in a large-scale rotating convection apparatus in a horizontal plane at mid-height to study the rich flow phenomenology of the geostrophic regime of rotating convection. We quantify the horizontal length scales of the flow using spatial correlations of vertical velocity and vertical vorticity, reproducing features of the convective Taylor columns and plumes flow states both part of the geostrophic regime. Additionally, we find in this horizontal plane an organisation into a quadrupolar vortex at higher Rayleigh numbers starting from the plumes state.}

\begin{document}

\maketitle

\section{Introduction}
Rayleigh--B\'enard convection, the flow in a fluid layer between two parallel horizontal plates where the bottom plate is at a higher temperature than the top, is a canonical model for buoyancy-driven flows. The addition of rotation is a popular and interesting extension, as the combination of buoyant forcing and rotation captures two of the principal constituents of many flows in geophysics and astrophysics. It is a mathematically well-defined problem that lends itself well to both numerical and experimental investigation (with appropriate lateral confinement) and stands out as a turbulent flow problem where numerical and experimental results can be compared one-to-one to great success \cite{scl13}.

Three principal parameters are required to describe rotating Rayleigh--B\'enard convection (RRBC). Here we shall use the Rayleigh number $Ra=g\alpha\Delta TH^3/(\nu\kappa)$, quantifying the strength of thermal forcing relative to dissipation, the Ekman number $E=\nu/(2\Omega H^2)$, the ratio of viscous forces to Coriolis forces, and the Prandtl number $Pr=\nu/\kappa$, describing the diffusive properties of the fluid. Here $g$ is the gravitational acceleration, $\Delta T$ the temperature difference between the plates and $H$ their vertical separation, $\Omega$ represents the rotation rate, and $\alpha$, $\nu$ and $\kappa$, respectively, are the coefficient of thermal expansion, kinematic viscosity and thermal diffusivity of the fluid. Another popular parameter is the convective Rossby number $Ro_c=E\sqrt{Ra/Pr}$ that combines the three previous parameters to directly compare the strength of buoyancy to Coriolis forces. The most popular geometry for experiments is an upright cylinder, where the diameter-to-height aspect ratio $\Gamma=D/H$ characterises its shape.

As a result of their vast proportions, geophysical and astrophysical flows are governed by extreme values of the dimensionless parameters defined before, with $Ra\gtrsim 10^{15}$ and $E\lesssim 10^{-10}$ \cite{ms99,m00,ss11,accjknss15}. At the same time, the critical Rayleigh number $Ra_C$ for onset of convective motion is significantly higher when rotation is applied: for $Pr>0.68$ linear stability theory gives the asymptotic result $Ra_C=8.6956E^{-4/3}$ in the limit of small $E$ \cite{c61}. So, despite the huge $Ra$ values most of these natural flows are still dominated by rotation, implying that the supercriticality $Ra/Ra_C$ retains more modest values. This rotational constraint leads to the dominant force balance being between pressure gradient and Coriolis force, the so-called geostrophic balance \cite{g68}. In recent years it was found that this geostrophic state of rotating convection displays a diverse set of flow structures (cells, convective Taylor columns (CTCs), plumes and geostrophic turbulence (GT)) \cite{jrgk12,k21}. Each of these flow states is expected to display characteristic scaling behaviour for flow statistics, like the convective heat transfer and intensity of velocity fluctuations. To enter this flow state one needs to resort to specialised tools: asymptotically reduced models (e.g. Refs. \cite{sjkw06,jrgk12}), large-scale numerical simulations on fine meshes (e.g. Refs. \cite{sljvcrka14,kopvl16,amcock20}) or large-scale experiments \cite{cajk18}. Next to heat transfer measurements, there is a great demand for experimental flow analysis of geostrophic convection; here we present the first results in that direction.

In this work we employ our large-scale convection apparatus TROCONVEX \cite{cajk18,cmak20} to obtain experimental flow data with extensive coverage of the geostrophic regime, achieving unprecedented values of the governing parameters. The measurement method is stereoscopic particle image velocimetry (SPIV) \cite{rwwk07}, a technique that resolves full three-component flow velocity vectors in a planar cross-section. We explore the various flow features that make up the geostrophic regime and quantify their characteristic horizontal length scales using spatial correlations of vertical velocity and vertical vorticity, testing recent scaling arguments \cite{gcs19,ahj20} for the horizontal length scale of rapidly rotating convection. This correlation procedure has been employed to interpret results of numerical simulations \cite{nrj14}. We have successfully applied this method to experimental velocimetry results from a smaller experiment \cite{rkc17}. We distinguish the different flow structures from these spatial autocorrelations and the related length scales. At the same time, we search for so-called large-scale vortices (LSVs): simulations have shown that the GT state can develop a striking upscale energy transfer leading to the growth of domain-filling LSVs \cite{jrgk12,fsp14,ghj14,kopvl16,amcock20}. LSVs are so far only observed in numerical simulations of RRBC without lateral confinement: depending on the fluid and the operating parameters, a single LSV or a large dipolar vortex could be observed \cite{sa18,amcock20}. It is thus not clear whether LSVs can manifest in RRBC experiments, too. Finally, with these flow measurements we explore the regime of rotation-influenced turbulence (RIT) that we inferred from our earlier heat flux and temperature measurements \cite{cmak20}, with largely unknown flow properties.

\section{Experimental arrangement}
TROCONVEX is a large-scale rotating convection apparatus using water as the working fluid. Here we provide a short description; for details we refer to our previous works \cite{cajk18,cmak20}. The convection cell is an upright cylinder of inner diameter $D=0.39\un{m}$. Its modular structure allows heights $H$ between $0.8$ and $4\un{m}$; here we use $H=2\un{m}$ exclusively for an aspect ratio $\Gamma\approx 1/5$. The top and bottom plates are made of copper. The bottom plate is electrically heated and regulated to a constant temperature $T_b$. The top plate is equipped with a double spiral groove through which cooling liquid is circulated; a thermostated bath and chiller combined regulate the temperature to a constant value $T_t$. This convection cell is placed on a rotating table.

\begin{figure}
\centerline{\includegraphics[width=0.262\textwidth]{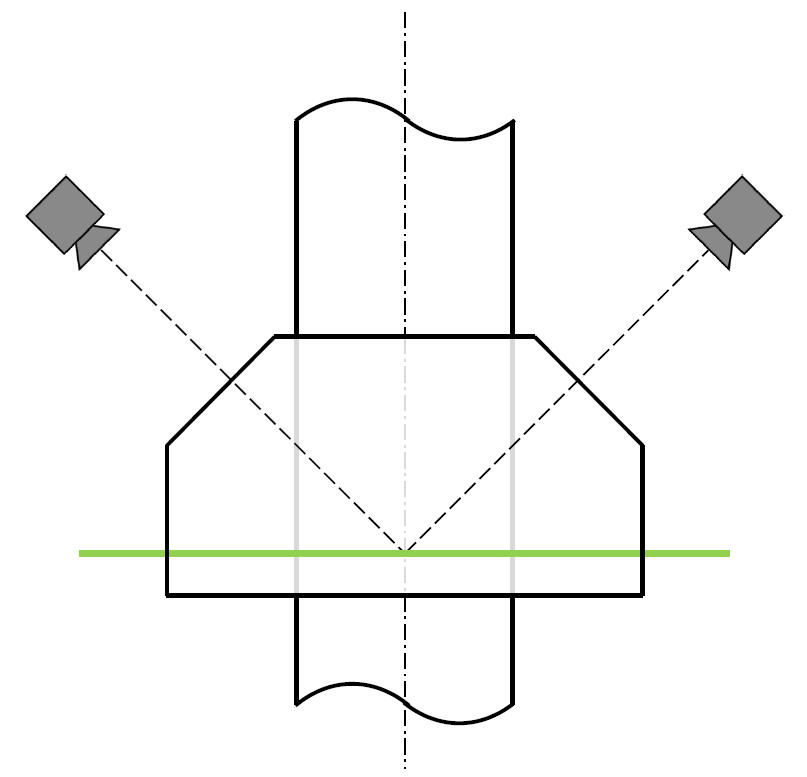}}
\caption{\label{fi:sketch_setup}Sketch of the SPIV arrangement. A water-filled acrylic prism surrounds the cylinder. The laser light sheet (green line) crosses the cylinder at mid-height. The $45^\circ$ oblique planes of the prism facilitate imaging of the intersection area of the laser sheet with the cylinder using the cameras.}
\end{figure}

Different from the previous heat transfer measurements \cite{cmak20}, we now employ a custom-made transparent acrylic cylinder section. A sketch of the SPIV arrangement is shown in figure \ref{fi:sketch_setup}. A custom-made water-filled prism around the cylinder enables optical access from outside without too much diffraction on the cylinder surface. A laser light sheet $\approx 3.5\un{mm}$ thick crosses the tank horizontally at mid-height, pulsing at frequencies of $7.5$ or $15\un{Hz}$; chosen according to the typical flow speeds. The water is seeded with polyamid seeding particles of nominal size $5\un{\upmu m}$. The illuminated particle images are recorded with two cameras (Jai SP-500M-CXP2; $5\un{Mpixel}$) placed on opposite sides of the cylinder at an angle of $\approx 45^\circ$ with the vertical. Scheimpflug adapters \cite{rwwk07} rotate lens and image plane so that the full intersection area is imaged in focus. This stereoscopic arrangement allows for SPIV evaluation \cite{rwwk07} of all three velocity components in the light sheet plane; the resulting velocity field fits $122$ vectors in the diameter at a vector separation of $3.2\un{mm}$ in both horizontal directions. Here we analyse between $3000$ and $9000$ vector fields per experiment, a duration of $200-600\un{s}$.

\begin{figure}
\includegraphics[width=0.48\textwidth]{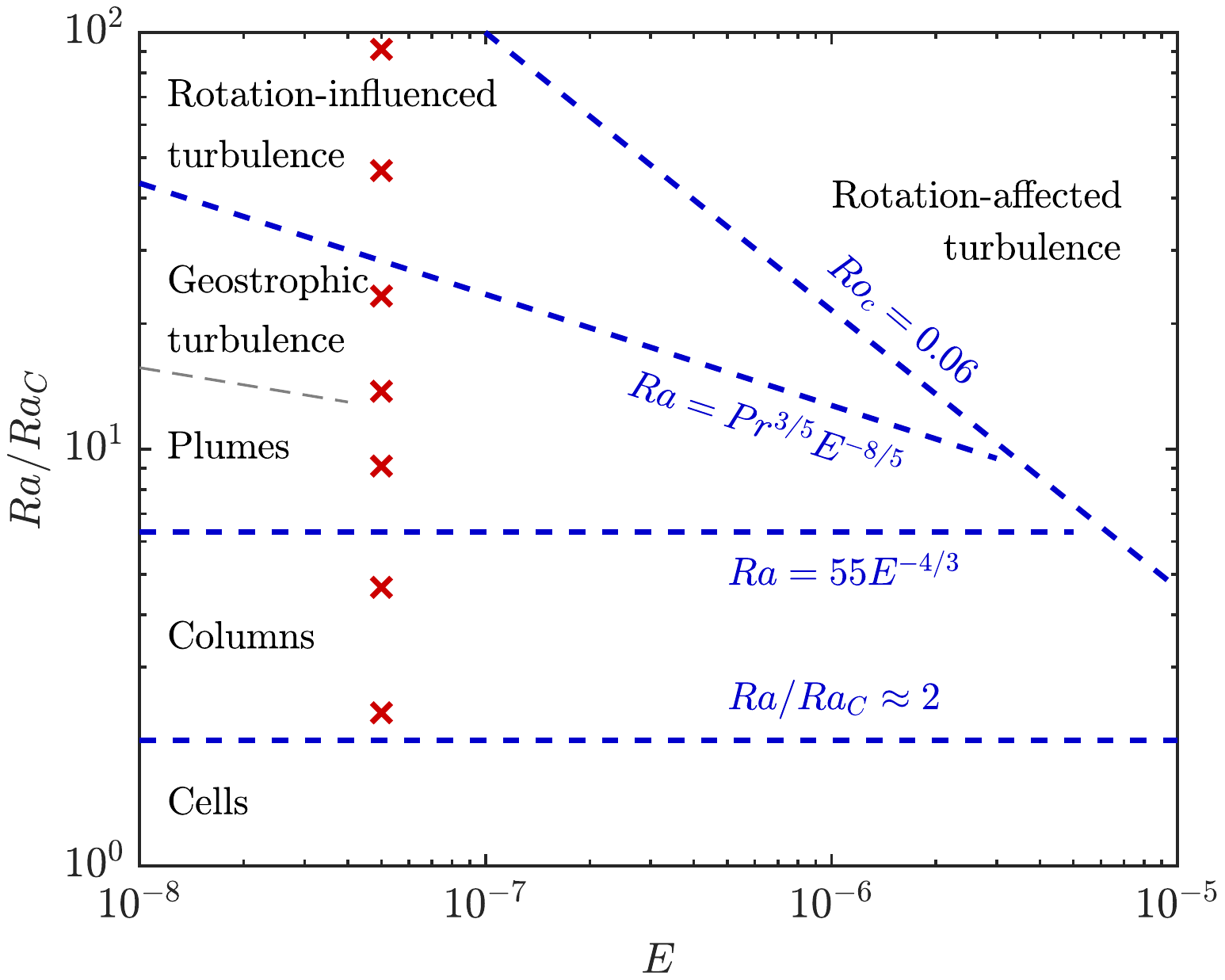}
\caption{\label{fi:phase_diag}Phase diagram for rotating convection in water. The expected flow structures and transitions between them (based on heat-flux and temperature measurements from TROCONVEX \cite{cmak20}) are labeled in the diagram. The experiments discussed in this work are indicated with red crosses.}
\end{figure}

We have selected seven operating conditions that capture various flow subregimes. These settings are indicated in a phase diagram in figure \ref{fi:phase_diag} with red crosses. The mean temperature $T_m=(T_b+T_t)/2$ is kept at $31\un{^\circ C}$, i.e. $Pr=5.2$. A constant rotation rate $\Omega=1.9\un{rad/s}$ is applied, i.e. $E=5\times 10^{-8}$. A major concern in experiments of this kind is the influence of centrifugal buoyancy. Centrifugal acceleration tilts the local vector of gravitational acceleration radially outward, away from the vertical, leading to possibly significant deviations from the intended flow configuration \cite{ha18,ha19}. The importance of centrifual acceleration is typically quantified as the Froude number $Fr=\Omega^2 D/(2g)$. For the experiments discussed here $Fr=0.07$; at this value we could not observe significant up/down asymmetry in our sidewall temperature measurements \cite{cmak20} pointing at negligibly small impact of centrifugal buoyancy.

We set bottom and top temperatures such that $Ra\in\{0.11, 0.22, 0.43, 0.65, 1.1, 2.2, 4.3\}\times 10^{12}$, or, correspondingly, $Ra/Ra_C\in\{2.3, 4.7, 9.1, 14, 23, 47, 91\}$. Note that it is hard in practice to get to smaller $Ra/Ra_C$ values in large setups like TROCONVEX, given that the small temperature differences required ($\Delta T < 0.5\un{^\circ C}$) cannot be controlled accurately enough at these scales \cite{cajk18,cmak20}. That is why we cannot enter the cellular state and only partially the CTC state. In addition, we consider one nonrotating case at $Ra=6.5\times 10^{11}$ for comparison.

\section{Spatial autocorrelation}
We can quantify and evaluate the spatial structure of the flow features by using spatial autocorrelations. This procedure has been applied to geostrophic convection results from simulations \cite{nrj14}, where the authors have shown that the autocorrelation nicely recovers the typical radial structure of flow features like cells and CTCs, and later also to experimental data based on PIV and 3D-PTV (three-dimensional particle tracking velocimetry) \cite{rkc17}. Here we define the spatial autocorrelation of a scalar variable $f$ as
\begin{equation}
R_f(r)=\frac{\left<f(\vect{x})f(\vect{x}+\vect{r})\right>_{A,t}}{\left<f^2(\vect{x})\right>_{A,t}} \, ,
\end{equation}
where $\vect{r}$ is the separation vector with length $r=|\vect{r}|$. The angular brackets $\left< \ldots \right>_{A,t}$ denote a spatial averaging over all available positions $\vect{x}$ within the cylinder cross-section $A$ and all possible $\vect{r}$ emanating from there, and we apply averaging in time $t$ as well. The effective area $A$ for the autocorrelation is deliberately reduced to exclude the wall mode that is found near the sidewall \cite{wamcck20,zghwzaewbs20,fk20,s20,zes21,wgbw21}; the thickness measure from \cite{wamcck20} is employed. Within this reduced area we assume horizontal homogeneity; $\vect{r}$ can be the vector connecting any two points within that area. We use zero padding and 2D fast Fourier transforms to calculate $R_f$, where the resulting (axisymmetric) 2D correlation graphs are converted to 1D using binning with circular shells of width $3.7\un{mm}$.

\begin{figure*}[!t]
\begin{tabular}{@{}lll@{}}
\includegraphics[scale=0.254]{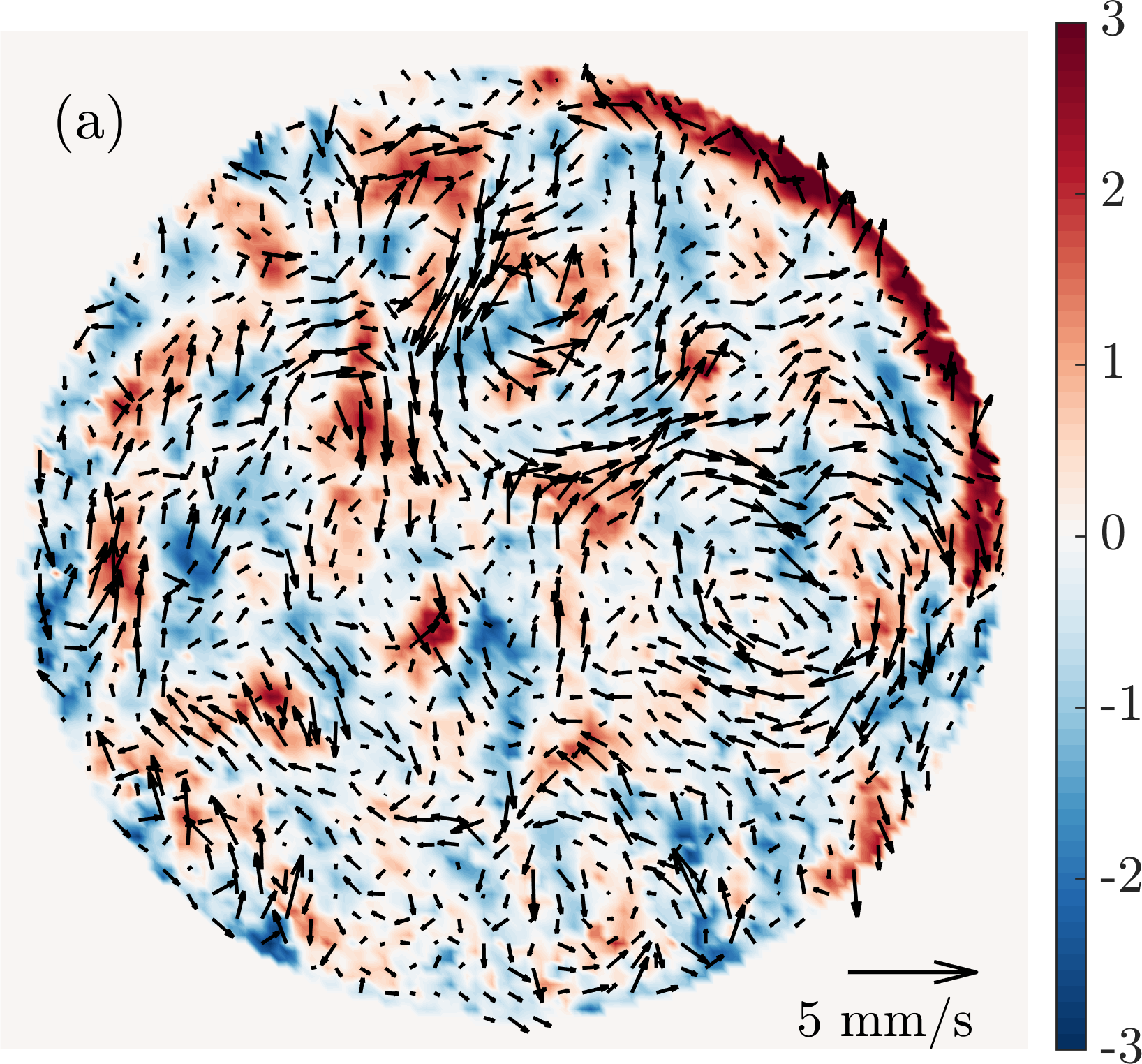} &
\includegraphics[scale=0.254]{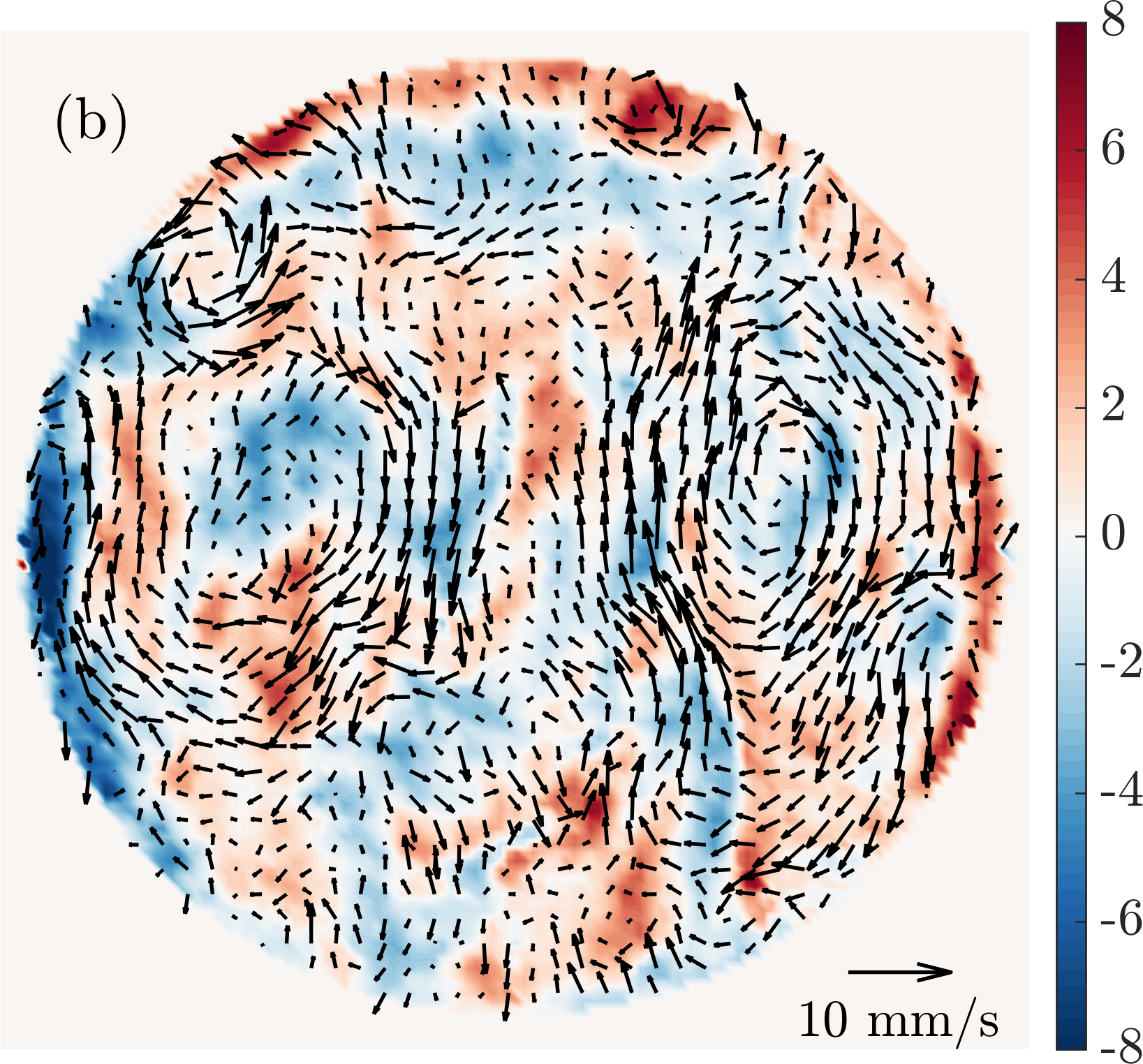} &
\includegraphics[scale=0.254]{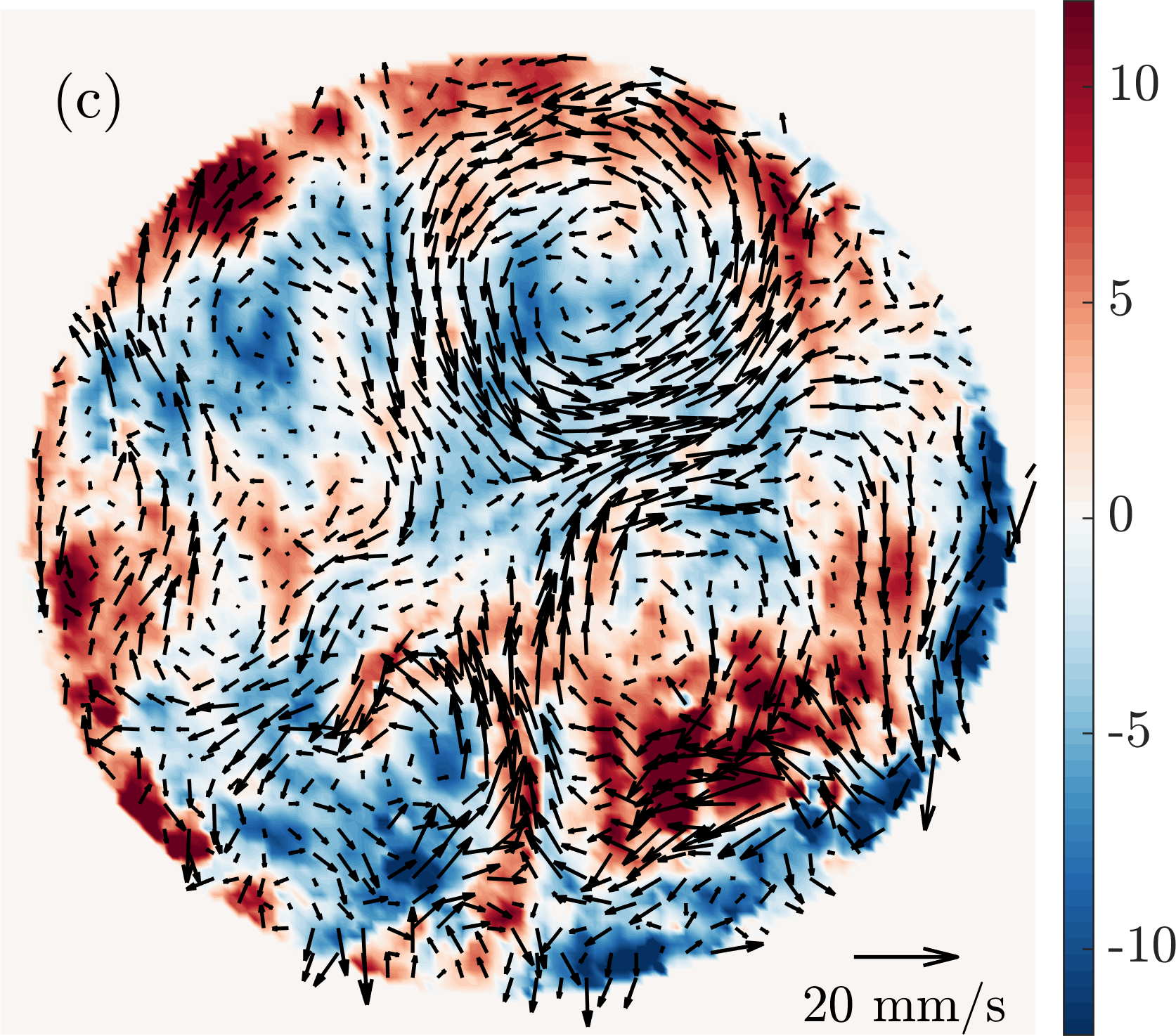} \\
\includegraphics[scale=0.254]{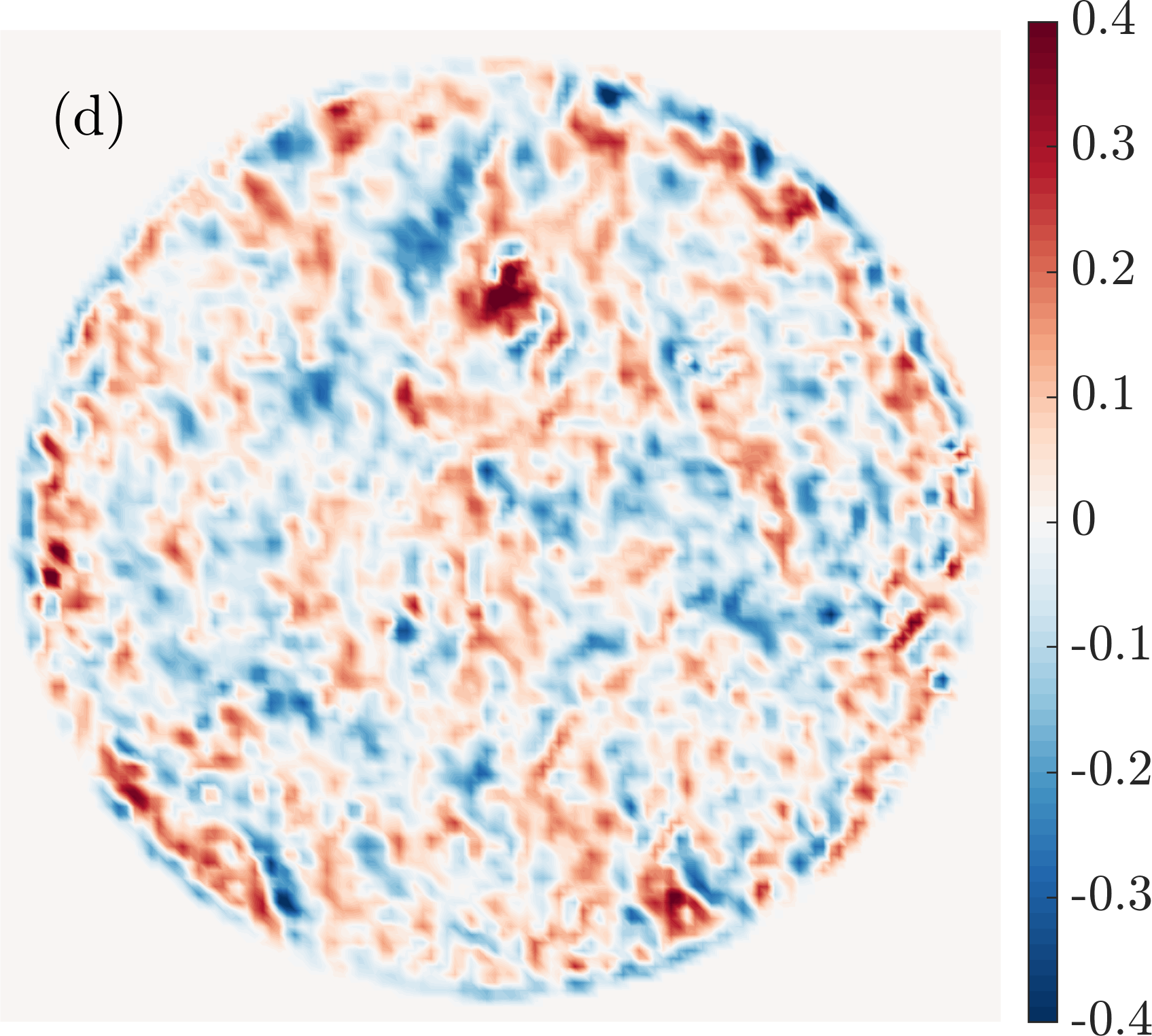} &
\includegraphics[scale=0.254]{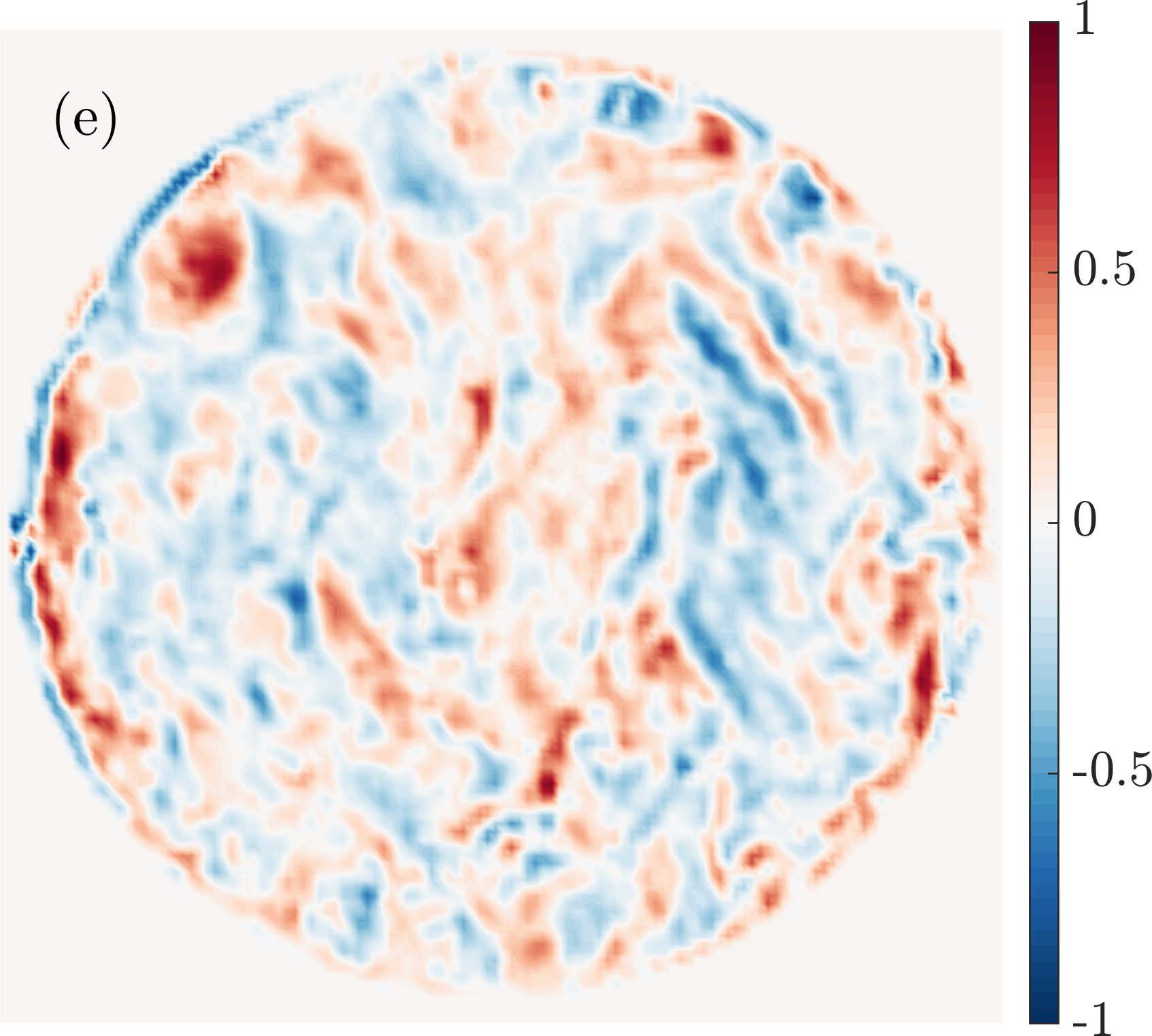} &
\includegraphics[scale=0.254]{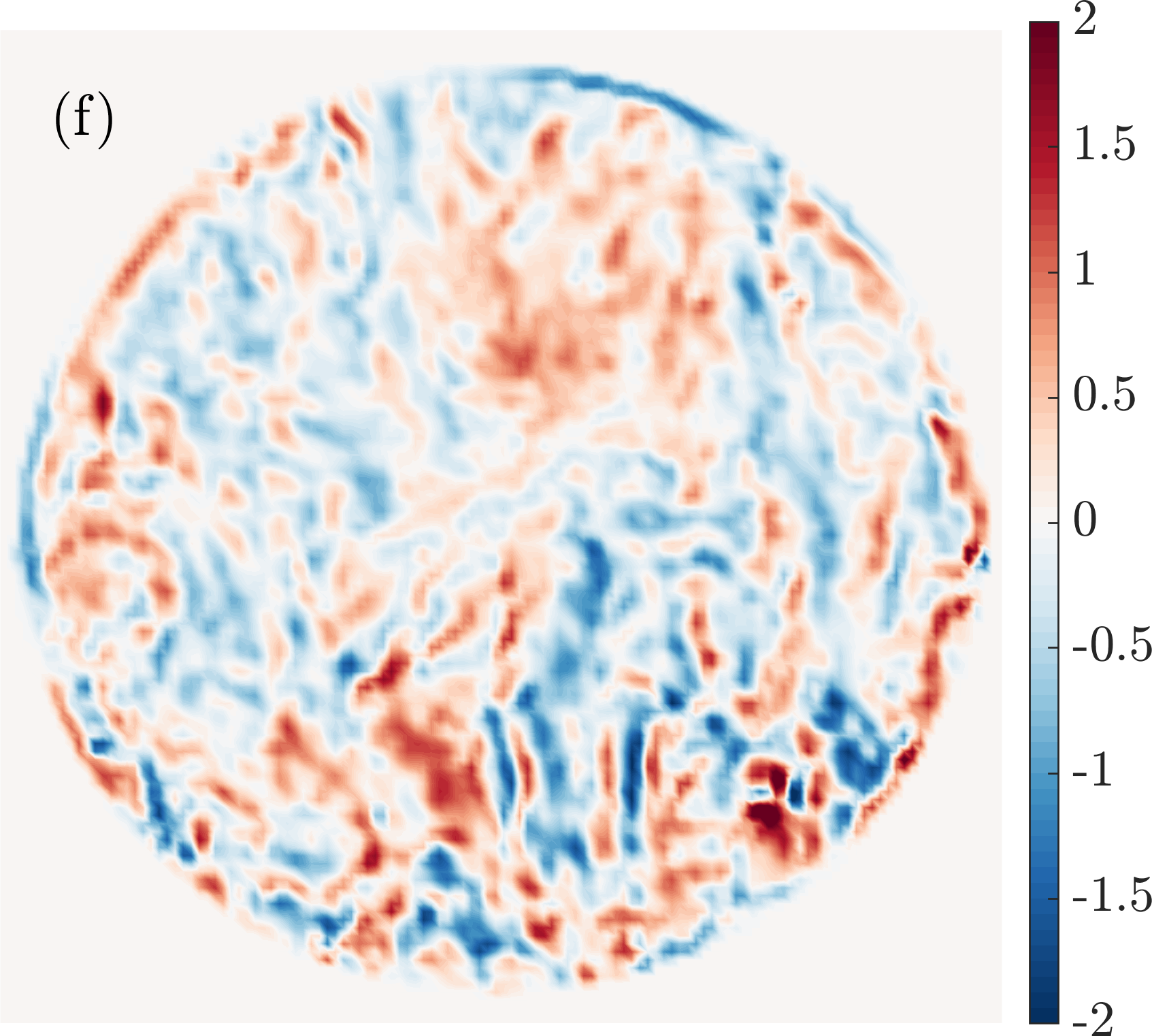}
\end{tabular}
\caption{\label{fi:snaps}(a--c) Flow velocity snapshots at mid-height for (a) $Ra/Ra_C=2.3$, (b) $Ra/Ra_C=14$, (c) $Ra/Ra_C=91$. The background colour indicates vertical velocity in $\un{mm/s}$; arrows represent the in-plane velocity components. Only one ninth of the total number of vectors is plotted for clarity of the images. (d--f) Snapshots of vorticity $\zeta$ in $\un{1/s}$ at mid-height for (d) $Ra/Ra_C=2.3$, (e) $Ra/Ra_C=14$, (f) $Ra/Ra_C=91$.}
\end{figure*}

Once the spatial autocorrelation $R_f(r)$ is known, we can compute the characteristic correlation length (or integral length scale \cite{p00}) as
\begin{equation}
L_f=\int_0^\infty R_f(r) \, \upd r \, .
\end{equation}

Nieves et al. \cite{nrj14} considered primarily correlations of temperature fluctuations $\theta$, but also of vertical velocity $w$ and vertical vorticity $\zeta=(\vect{\nabla}\vect{\times}\vect{u})_z$, where $\vect{u}=(u,v,w)$ is the velocity vector. Rajaei et al. \cite{rkc17} could only consider correlation of $\zeta$ from their regular PIV measurements. Here we shall consider both $w$ and $\zeta$. We expect from theoretical models describing the cellular and CTC states as single-wavenumber modes \cite{pkhm08,gjwk10} that $\theta$, $w$ and $\zeta$ lead to very similar correlation graphs. For plumes and GT (and possibly RIT) the correlations for $w$ and $\zeta$ may develop significant differences in correlation length \cite{nrj14}.

\section{Flow snapshots}
To get a first impression of the flow field that develops, we present instantaneous snapshots of velocity and vorticity at three different $Ra/Ra_C$ values in figure \ref{fi:snaps}. Panels (a,d) are in the range where CTCs are expected. We do observe vortices present in this flow, though the shields are hard to distinguish. One contributing factor is that we measure at mid-height where the columns are expected to be weaker; their vertical structure is torsional, with positive (cyclonic) vertical vorticity at one vertical end and negative (anticyclonic) vorticity at the other end \cite{s97,pkhm08,gjwk10}. Here we observe in animations of the flow (see the supplementary material) that the flow state is quite dynamical; no quasi-steady vortex grid is formed. Instead, the vortices wander around and can interact with other vortices and the wall mode, making them not as long-lived. Near the sidewall, the signature of the wall mode \cite{wamcck20,zghwzaewbs20,fk20,s20,zes21,wgbw21} can be seen: the top right half of the circumference displays a prominent vertically upward flow (red colour), while the opposite side shows downward flow (blue colour; though not as prominent in this snapshot). A prominent dynamical feature of this wall mode is the presence of two jets emanating from the azimuthal positions where the up- and downward lobes meet pointing radially inward \cite{wamcck20}. These jets, more easily identified in animations of the flow (see supplementary material), appear to play an important role in setting the flow structures in the bulk in motion. Similar structures have been found in magnetoconvection simulations \cite{lks18,azks20}.

At the higher $Ra/Ra_C=14$ (figure \ref{fi:snaps}b,e), representative of plumes, a qualitatively similar flow field is observed. Velocities are larger given that the thermal forcing is stronger. It is readily observed that the vertical velocity is partitioned into larger patches, i.e. a larger correlation length is anticipated. At the same time, the vorticity remains confined to narrow patches. The wall mode is still quite prominently visible, though the fluctuations in the bulk have increased in magnitude compared to the amplitude of the wall mode \cite{wamcck20} so that its relative dynamical significance is diminished.

Figures \ref{fi:snaps}(c,f) display a snapshot at the highest $Ra/Ra_C=91$ considered here, in the RIT range. Velocities are even larger than in the previous panel due to stronger thermal forcing. The correlation length of vertical velocity has become even larger. Contrarily, the vorticity field still reveals finer scales comparable to the previously discussed cases (figure \ref{fi:snaps}d,e).

\begin{figure}
\centerline{\includegraphics[scale=0.254]{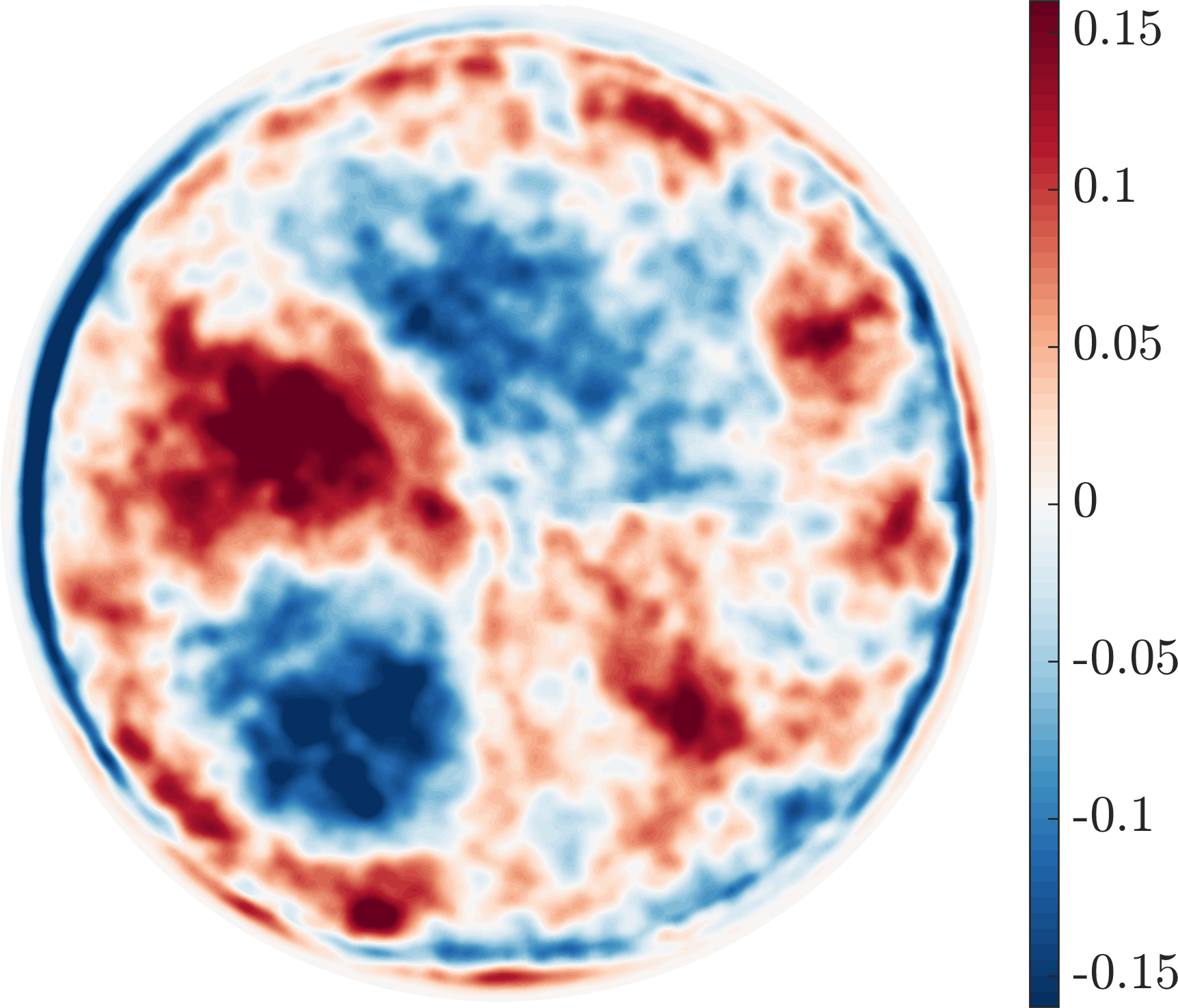}}
\caption{\label{fi:qpv}Orientation-compensated mean vorticity field (in $\un{1/s}$) at $Ra/Ra_C=47$.}
\end{figure}

An interesting new feature --- that can be recognised from animations of the flow field as a function of time (see supplementary material) --- is the organisation into a quadrupolar vortex state that fills the cross-sectional area with two cyclonic and two anticyclonic swirls. This flow arrangement forms a secondary large-scale flow somewhat reminiscent of LSV dynamics as previously observed in simulations in laterally unbounded domains. The wall mode couples to this quadrupolar vortex state (or the other way around) to organise the entire flow field into a pattern that displays a slow azimuthal drift \cite{wamcck20}. This drift makes simple time-averaging ineffective to further illustrate this structure. Instead, we employ an orientation averaging similar to Ref. \cite{wamcck20}. We determine the phase angle of the wall mode at each time step by tracking the precession of the vertical velocity signal on a circular trajectory close to the sidewall. Each snapshot is rotated by its corresponding phase angle so that the orientation matches for each snapshot, then the velocity field is averaged to get a clear view of the mean flow profile. Figure \ref{fi:qpv} shows the result of such averaging, an orientation-compensated mean vorticity field for the experiment at $Ra/Ra_C=47$. There is an organisation as a quadrupolar vortex with four cores of vorticity that fill the central part of the domain. We are currently considering this overall flow structuring, its origin and any relation with the wall mode in more detail.

For the nonrotating reference case (animation in the supplementary material) the well-known large-scale circulation (LSC \cite{agl09}; not to be confused with LSV) of nonrotating confined convection can be observed: a principal partitioning of vertical velocity into two patches (one with positive and one with negative vertical velocity) that cover the entire cross-section of the cylinder. Therefore, the largest correlation length is expected in this case, of the order of the radius of the cylinder.

\section{Results: autocorrelations}
\begin{figure}
\includegraphics[width=0.48\textwidth]{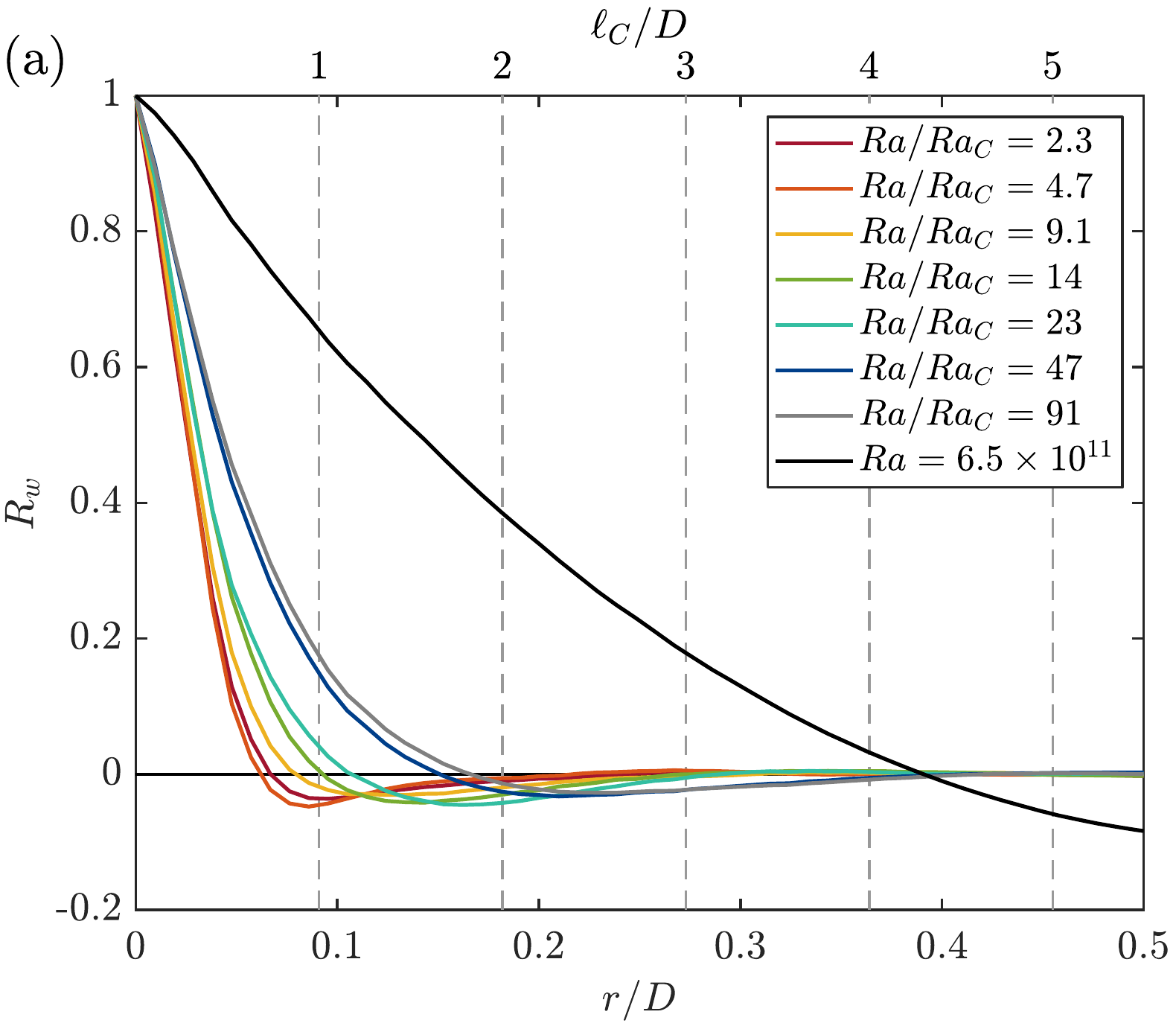}\\
\includegraphics[width=0.48\textwidth]{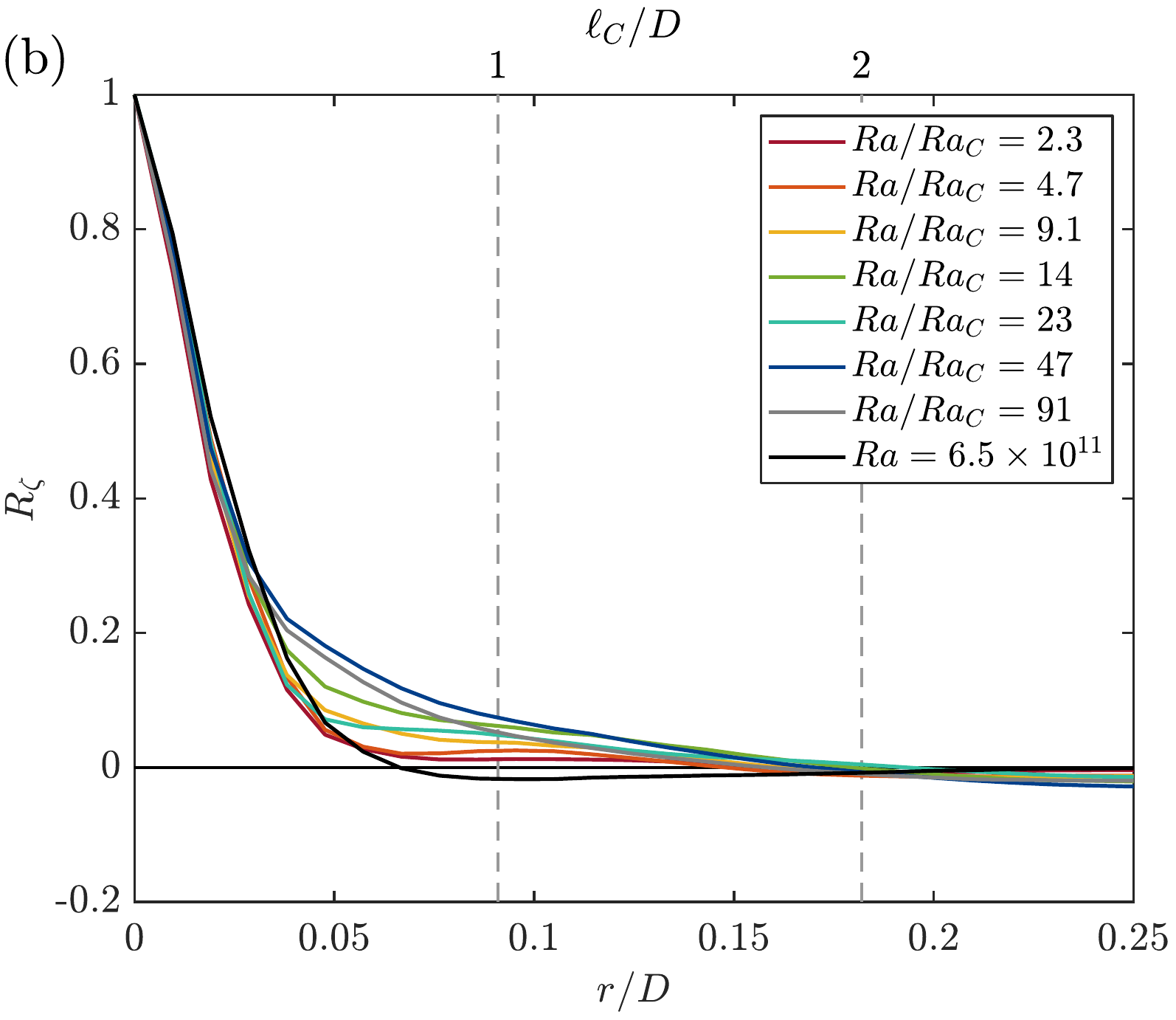}
\caption{\label{fi:corr}Spatial autocorrelations of (a) vertical velocity $w$ and (b) vertical vorticity $\zeta$. Length scales are normalised using the cell diameter $D$. Vertical dashed lines display multiples of $\ell_C$ for reference. Note the different horizontal axis ranges.}
\end{figure}

We plot spatial autocorrelation graphs for the rotating convection cases and the nonrotating reference case in figure \ref{fi:corr}. In these plots we also indicate units of $\ell_C$, the wavelength of convective instability ($\ell_C=4.8154E^{1/3}H$ for $Pr>0.68$ \cite{c61}) that plays an important role throughout the geostrophic regime as a characteristic horizontal scale \cite{sjkw06,jrgk12}. For $R_w$, the autocorrelation of vertical velocity, we see that the correlation is reduced as $r$ increases, reaches a negative minimum, then approaches zero at large $r$. As $Ra/Ra_C$ increases, stronger thermal forcing and more vigorous turbulence, the correlation length increases considerably. This is true even more so for the nonrotating case, where the minimum is out of the plotting range at $r/D=0.57$. In that case the correlation length is determined by the presence of the LSC; it is of the order of the cylinder radius. The $R_w$ results reported in Ref. \cite{nrj14} display a similar elongation of the correlation as $Ra/Ra_C$ increases. However, the regular oscillatory behaviour for the CTC flow range is not reproduced here. We do not see the development of a true ensemble of shielded Taylor columns. Instead, the bulk flow structures are moving around in a secondary circulation that appears to be set up by the strong wall mode in this flow range, with jets penetrating the bulk from laterally opposite sides of the cylinder cross-section \cite{wamcck20}. This prevents the formation and preservation of a quasi-steady grid of CTCs as observed in simulations on horizontally periodic domains, e.g. \cite{jrgk12,sljvcrka14}, as well as in smaller-scale experiments at larger $E$ and $\Gamma$\cite{rkc17}. 

The vorticity correlations $R_\zeta$ (figure \ref{fi:corr}b) decay on comparatively smaller scales than $R_w$, in agreement with the expectations based on the snapshots of figure \ref{fi:snaps}. The initial decrease at small $r$ is quite similar for all the considered cases. This is in line with a gradual transition from the structures of convection at small supercriticality (e.g. the single-mode solutions \cite{pkhm08,gjwk10}), where equal scales are found for velocity and vorticity, to a situation in line with nonrotating 3D turbulence composed of thin vortex tubes (e.g. Ref. \cite{igk09}) with the vorticity correlation length roughly equal to the Kolmogorov length while the velocity correlation scale is the largest length scale in the flow. Then, the cases at the smallest values of $Ra/Ra_C$ display a small oscillation that was expected for the CTC state based on earlier simulations \cite{nrj14} and experiments \cite{rkc17}. Its wavelength corresponds quite nicely with the convective wavelength $\ell_C$. The oscillation is not as pronounced here as in the other studies, presumably due to the inability to form a quasi-steady CTC grid. For $Ra/Ra_C=9.1$ and higher (excluding the nonrotating case) we observe the occurrence of a larger length scale on which some correlation can be seen: starting from $r/D\approx 0.05$ these curves display a shallow downward slope, with a zero crossing at $r/D\approx 0.2$. The curves each reach a shallow minimum at $r/D\approx 0.26-0.30$ then asymptote to zero at large $r$. This correlation signature is a second indicator of the organisation into a quadrupolar vortex consisting of two cyclonic and two anticyclonic cells.

\begin{figure}
\includegraphics[width=0.48\textwidth]{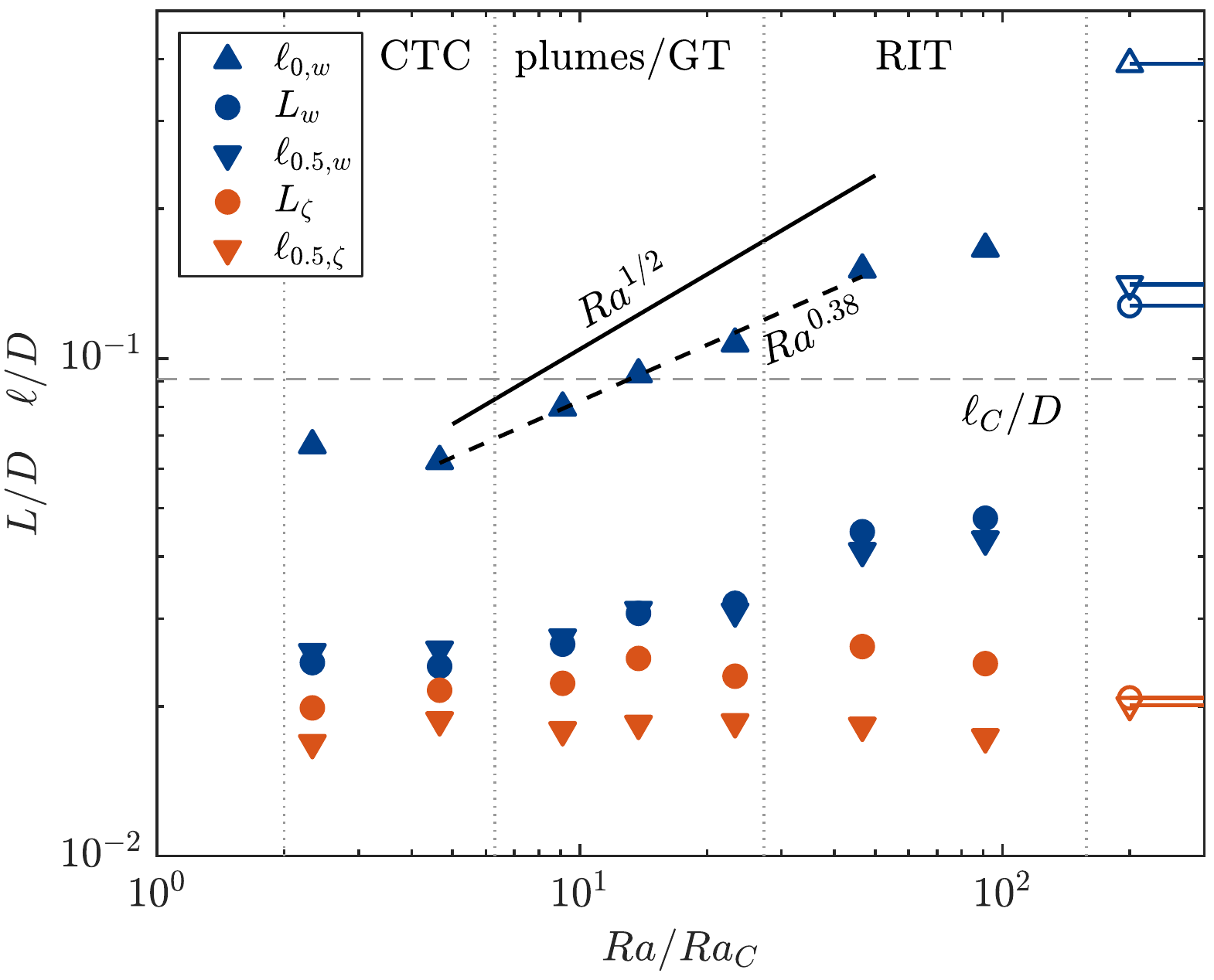}
\caption{\label{fi:lengthscales}Correlation length scales as a function of $Ra/Ra_C$. The line segments with open symbols are the corresponding results for the nonrotating reference case (same symbols and colours; not to scale in terms of $Ra/Ra_C$). Error intervals (not plotted) are equal to the symbol size or smaller. The horizontal dashed line indicates the convective wavelength $\ell_C$. Vertical dotted lines indicate the regime transitions of figure \ref{fi:phase_diag}. The black solid line indicates the scaling $\ell\sim Ra^{1/2}$; the black dashed line is a fit $\ell\sim Ra^{0.38}$ to the data.}
\end{figure}

We can further quantify and compare these autocorrelation results by deriving characteristic length scales from them. We consider the integral scales $L_w$ and $L_\zeta$ as defined before. Additionally, we define length scales based on the correlation magnitude: $\ell_{0.5,w}$ and $\ell_{0.5,\zeta}$ for the $r$ where the corresponding autocorrelation has the value $0.5$, and $\ell_{0,w}$ where $R_w$ crosses zero for the first time. The zero crossing is not as informative for $R_\zeta$ given the longer positive correlation that is observed. These length scales are plotted as a function of $Ra/Ra_C$ in figure \ref{fi:lengthscales}. It is clear that the vorticity-based scales are always smaller than their velocity-based counterparts. While $L_\zeta$ shows some variation with $Ra/Ra_C$, the smaller $\ell_{0.5,\zeta}$ remains more or less constant throughout. Indeed, the initial decay of $R_\zeta$ is quite similar in all cases. At the two smallest $Ra/Ra_C$ values considered here the velocity-based scales $L_w$ and $\ell_{0.5,w}$ are of comparable size to their vorticity-based counterparts, as expected based on prior results for the CTC state \cite{pkhm08,gjwk10,nrj14,rkc17}. For larger $Ra/Ra_C\gtrsim 9$ the velocity-based scales become increasingly larger. Based on this observation we expect that the CTC-to-plumes transition takes place between $Ra/Ra_C=4.7$ and $9.1$. This is in agreement with the reported transition $RaE^{4/3}=55$ (or $Ra/Ra_C=6.3$) for the asymptotic simulations and fully in line with conclusions based on our earlier heat-flux and temperature measurements in the same setup \cite{cmak20}. Beyond that transition, in the plumes state, the velocity correlation widens, though for the highest two $Ra/Ra_C$ values some saturation can be observed. The correlation graphs for these cases, in the RIT range, are in line with the plumes cases in terms of shape. The saturation of $\ell_{0,w}$ at such length is in line with the organisation into a quadrupolar structure, where correlation up to about one fourth of the diameter is expected. Note the significant difference with the nonrotating case, where correlation continues up to about half the diameter due to presence of the LSC with the cross-sectional area divided into one half upward and one half downward flow.

Two recent works have considered the horizontal length scale of convection in the geostrophic regime. Guervilly et al. \cite{gcs19} combine results of various numerical models to find an effective scaling $\ell\sim Ro_U^{1/2}\sim (RaE^2/Pr)^{1/2}$ with the Rossby number $Ro_U$ based on a measured velocity scale $U$. In our notation this amounts to $\ell\sim Ro_c$. They only find this scaling at very small $E\lesssim 10^{-9}$. Aurnou et al. \cite{ahj20} provide theoretical scaling arguments based on the so-called CIA (Coriolis--Inertial--Archimedean) force balance that also predict $\ell\sim Ro_c$. In our experiments at constant $E$ with variation of $Ra$ this translates to $\ell\sim Ra^{1/2}$. This scaling slope is included in figure \ref{fi:lengthscales} with the solid black line; a trend clearly steeper than our data. A power law fit to our data for $4.7\le Ra/Ra_C\le 47$ (dashed black line) renders a scaling $\ell\sim Ra^{0.38}$. Looking at figure 4(b) of Guervilly et al. \cite{gcs19}, our shallower scaling corresponds nicely to the shallower trend of their data for $10^{-9}\lesssim E\lesssim 10^{-7}$, which indeed encloses our $E$ value. While the scaling of the length scale is similar, comparison of the magnitude is not possible due to differences in domain (sphere vs. cylinder) and $Pr$ value (0.01 vs. 5.2).

\section{Conclusion}
We have performed stereoscopic particle image velocimetry measurements in rotating thermal convection in the geostrophic regime at small Ekman number $E=5\times 10^{-8}$ where the effects of rotation are prominent. The flow phenomenology has been quantitatively analysed using spatial correlations of vertical velocity and vertical vorticity. The correlation length scales based on vertical vorticity remain reasonably constant over the considered range of supercriticality values $2.3\le Ra/Ra_C\le 91$, in line with observations that the critical wavelength $\ell_C$ for onset of convection is an important horizontal length scale throughout the geostrophic regime. Correlation length scales of vertical velocity grow with increasing $Ra/Ra_C$ and can be used to identify different flow states: the state of convective Taylor columns (CTC) for $Ra/Ra_C\lesssim 6$, the plumes state for $Ra/Ra_C\gtrsim 6$ changing gradually into the previously uncharacterised state of rotation-influenced turbulence (RIT), in good agreement with earlier results from reduced numerical models in the same $E$ range \cite{gcs19}. The plumes/GT and RIT ranges display an interesting new organisation into a quadrupolar vortex (recognised from orientation-compensated mean vorticity fields) which is presumably preferred over a single LSV or a dipole \cite{sa18,amcock20} in this confined domain.

Throughout these measurements we can identify the wall mode, a coherent vertical flow structure near the sidewall consisting of one lobe with upward transport on one half of the circumference and one downward lobe on the other half. The prominence of the wall mode relative to the fluctuations in the interior diminishes at higher $Ra/Ra_C$. At low $Ra/Ra_C$ the jets emanating from the wall mode set the vortical structures in horizontal motion; contrary to the results from horizontally periodic simulations no quasi-steady CTC grid forms. The nature and origin of the coupling of the wall mode and the interior flow, at both high and low limits of $Ra/Ra_C$, is an open question that we want to address later.

The study of rotating convection in the geostrophic regime poses challenges to experimentalists and numericists alike. Nonetheless, recent numerical works have elucidated the rich flow phenomenology that develops in this convection setting. Here, we contribute experimental results at more extreme values of the governing parameters that expand our understanding of this intriguing and geo-/astrophysically relevant flow problem, despite the influence of unexpected yet intriguing features like the persistent wall mode and its interaction with the bulk flow.

\acknowledgments
M.M., A.J.A.G. and R.P.J.K. received funding from the European Research Council (ERC) under the European Union's Horizon 2020 research and innovation programme (Grant agreement No. 678634).

\bibliographystyle{eplbib}
\bibliography{kunnen}

\end{document}